\documentclass[]{elsart}

\usepackage{amssymb}

\newcommand{\eeq}{\end{equation}}
\newcommand{\beq}{\begin{equation}}
\newcommand\slabel[1]{\label{#1}}

\newcommand{\AmS}{{\protect\the\textfont2
  A\kern-.1667em\lower.5ex\hbox{M}\kern-.125emS}}

\begin{document}

\begin{frontmatter}

% Title, authors and addresses

% use the thanksref command within \title, \author or \address for footnotes;
% use the corauthref command within \author for corresponding author footnotes;
% use the ead command for the email address,
% and the form \ead[url] for the home page:
% \title{Title\thanksref{label1}}
% \thanks[label1]{}
% \author{Name\corauthref{cor1}\thanksref{label2}}
% \ead{email address}
% \ead[url]{home page}
% \thanks[label2]{}
% \corauth[cor1]{}
% \address{Address\thanksref{label3}}
% \thanks[label3]{}

\title{Classification of Cosmological Trajectories}

% use optional labels to link authors explicitly to addresses:
% \author[label1,label2]{}
% \address[label1]{}
% \address[label2]{}

\author{Qi Guo\thanksref{gq}}\author{,  Rong-Gen Cai\thanksref{rgc}}

\thanks[gq]{guoqi@itp.ac.cn}
\thanks[rgc]{cairg@itp.ac.cn}

\address{Institute of Theoretical Physics,
Chinese Academy of Sciences,
 P.O.Box 2735, Beijing 100080,
China}

\begin{abstract}
In the context of effective Friedmann equation we classify the
cosmologies in multi-scalar models with an arbitrary scalar
potential $V$ according to their geometric properties. It is shown
that all flat cosmologies are geodesics with respect to a
conformally rescaled metric on the `augmented' target space.
Non-flat cosmologies with $V=0$ are also investigated. It is shown
that geodesics in a `doubly-augmented' target space yield
cosmological trajectories for any curvature $k$ when projected
onto a given hypersurface.
\end{abstract}

\end{frontmatter}

% main text

\section{Introduction}
Scalar fields have come to play a central role in modern
cosmology. Current astronomical observations indicate that our
universe is undergoing an acceleration sourced by dark energy
\cite{observation}. As prospective dark energy candidates, a wild
variety of scalar field dark energy models have recently been
proposed,including quintessence \cite{que}, k-essence
\cite{kess}\cite{kkess}, and ghost(phantom)\cite{phantom}.
However, the analysis on properties of dark energy from the recent
observations mildly favor models with state equation parameter
$\omega$ crossing -1 in the near
past\cite{phantom}\cite{doublescalar} which could not be realized
only by a single scalar\cite{nogo}. Consequently, study of
multi-scalar fields becomes one of the most intriguing
subjects\cite{gr}.

On the other hand, the Friedmann equation forms the starting point
for almost all investigations in cosmology. Over the past few
years possible corrections to the Friedmann equation have been
derived or proposed in a number of different contexts, generally
inspired by braneworld investigation\cite{effF}\cite{effF2}. These
modification are often of a form that involves the total energy
density $\rho$. In\cite{gr}, multi-scalar coupled to gravity is
studied in the context of conventional Friedmann cosmology. It is
found that the cosmological trajectories can be viewed as geodesic
motion in an `augmented' target space.In this paper, taking the
$N$ scalar fields as coordinates of a Riemannian target space with
metric $G_{ij}$ we present a unified framework to investigate the
geometric properties in a general cosmological background
characterized by(6). The General Relativity,
Randall-Sundrum\cite{RS1}, and Gauss-Bonnet\cite{GB} scenarios
correspond to $n=1$,$n=2$ and $n=2/3$,respectively.

We start the formalism in the flat universe in the context of
modified Friedmann equation. It is shown that solutions of gravity
coupled to $N$ scalar field with $V=0$ can be viewed as null
geodesics in an `augmented' target space of dimension $N+1$ with a
Lorentzian signature metric. However, for $V\neq0$, the
cosmological trajectories are geodesics only with respect to a
conformally rescaled metric; the geodesic is timelike if $V>0$ and
spacelike if $V<0$. Investigation of purely kinetic energy driving
cosmology is of interest in its own right.For example, purely
kinetic k-essence can serve as a unified model for dark matter and
dark energy\cite{kkess}. Alternative time-coordinate is introduced
in this case. It turns out that cosmological solutions with pure
kinetic term is again null geodesic when $k=0$. For non-flat
cosmologies, the trajectories in `augmented' target space are
neither geodesic nor with global consistent causal property.
However, they can be viewed as projections of geodesics in a
`doubly-augmented' target space.

The organization of the paper is as follows. In the next section
we present the multi-scalar solutions in the context of modified
Freidmann equation. In Sec. 3 and 4 we develop the interpretation
of multi-scalar cosmologies as geodesic motion with two different
choice of metric. In Sec. 5 we introduce a light cone in which the
universe undergoes an acceleration. This paper is ended in Sec. 6
with some conclusions.

\section{Equations of Motion }
\label{sepunisec} We start by reviewing the equations of motion
for multi-scalar. We shall study cosmologies in
Friedmann-Robertson-Walker(FRW) background spacetime:

\beq \emph{d}s^{2}=-\emph{d}t^{2}+S(t)^{2}\emph{d}\Sigma_{k}^{2},
\eeq where the function $S(t)$ is the scale factor, and
$\Sigma_{k}^{2}$ represents the 3-dimensional spatial sections of
constant curvature $k$. Here $k$ is normalized to take values
$0,\pm1$.

The Lagrangian density for multi-scalar is

\beq L_{\phi}=
-\frac{1}{2}g^{\mu\nu}G_{ij}\partial_{\mu}\phi^{i}\partial_{\nu}\phi^{j}-V(\phi).\eeq
To simplify the equations we define
 \beq
(D^{2}_{t}\phi)^{i}:=\partial_{t}^{2}\phi^{i}+\Gamma^{i}_{
jk}\partial_{t}\phi^{j}\partial_{t}\phi^{k},
 \eeq
where $\Gamma^{i}_{jk}$ is the Levi-Civita connection for the
target space metric $G$. Then we obtain the equations of motion:

\beq D_{t}^{2}\phi^{i}+3H\partial_{t} \phi^{i}-2Va^{i}=0, \eeq

\beq a_{i}=-\frac{1}{2}\frac{\partial ln
|V|}{\partial\phi^{i}},\eeq

where $H=\partial_{t}S/S$ is the Hubble function. We consider an
effective Fridmann equation which is given by

\beq H^{2}=A\rho^{n}-k/S^{2} \slabel{friedmann}, \eeq where $A$
and $n$ are constants. $\rho$ is twice the energy density of
scalar fields:

\beq \rho=|\partial_{t}\phi|^{2}+2V, \eeq where
$|\partial_{t}\phi|^{2}$ is induced by the target space metric:
$|\partial_{t}\phi|^{2}=G_{\alpha\beta}\partial_{t}\phi^{\alpha}\partial_{t}\phi^{\beta}.$
Substituting Eqs.(4) to (6), we get the acceleration equation:

\beq
\partial_{t}^{2}S=AS\rho^{n-1}[2V-(3n-1)|\partial_{t}\phi|^{2}].
\eeq Set

\beq S(t)=e^{\beta(t)}.\eeq The Friedmann equation can be written
as

\beq
(\partial_{t}\beta)^{2}=A(|\partial_{t}\phi|^{2}+2V)^{n}-ke^{-2\beta},\eeq
the scalar field equation as

\beq
D_{t}^{2}\phi^{i}+3(\partial_{t}\beta)\partial_{t}\phi^{i}=0,\eeq
and the acceleration equation as

\beq
\partial_{t}^{2}\beta+\partial_{t}\beta=-A(|\partial_{t}\phi|^{2}+2V)^{n-1}[2V-(3n-1)|\partial_{t}\phi|^{2}].
\eeq

\section{Flat Cosmologies }
In this section we develop a geometric method to describe the
generic evolution of flat multi-scalar cosmologies. Define a new
variable $\gamma$ by

\beq
\partial_{t}\gamma=(\frac{\partial_{t}\beta}{\sqrt{A}})^{\frac{1}{n}}.
\eeq We may consider the (N+1) variables

\beq \Phi^{\mu}=(\gamma,\phi^{i}) \eeq as the elements
constructing an `augmented' target space. In this notion, the
Friedmann equation can be written as

\beq G_{\mu\nu}\partial_{t}\Phi^{\mu}\partial_{t}\Phi^{\nu}=-2V,
\eeq where

\beq
G_{\mu\nu}\partial_{t}\Phi^{\mu}\partial_{t}\Phi^{\nu}=-d\gamma^{2}+G_{ij}d\phi^{i}d\phi^{j}\eeq
is a $Lorentz-signature$ metric on the space. Therefore, the
cosmological trajectory in the augmented target space can be
classified by the signature of the potential $V$. It is null for
$V=0$, timelike for $V>0$ and spacelike for $V<0$.

Rewrite the acceleration equation (12) and scaler field equation
(11) with $\gamma$ instead of $\beta$:

\beq
\partial_{t}^{2}\gamma+3\sqrt{A}(\partial_{t}\gamma)^{n-1}|\partial_{t}\phi|^{2}=0,
\eeq

\beq
D_{t}^{2}\phi^{i}+3\sqrt{A}(\partial_{t}\gamma)^{n-1}\partial_{t}\alpha\partial_{t}\phi^{i}+a^{i}|\partial_{t}\Phi|^2=0,
\eeq where $|\partial_{t}\Phi|^{2}$ is induced by the augmented
target space metric:
$|\partial_{t}\Phi|^{2}=G_{\alpha\beta}\partial_{t}\Phi^{\alpha}\partial_{t}\Phi^{\beta}.$
We will find that after some special conformal transformation, all
cosmological trajectories are geodesics. Set conformal factor as

\begin{equation}\label{conformalfac}
\Omega=\cases{{1\over\sqrt{2}}\sqrt{\left|V\right|} \, e^{f(
\gamma)/2} & $V\ne0$ \cr {1\over\sqrt{2}}\ \, e^{f(\gamma)/2} &
$V=0$\,.}
\end{equation}
Denote the new metric by $\widetilde{G}$, satisfying

\beq \widetilde{G}_{\mu\nu}=\Omega^{2}G_{\mu\nu}.\eeq One can find

\begin{eqnarray}
\widetilde{D}^{2}_{t}\Phi^{\mu}\nonumber &\equiv&
\partial_{t}^{2}\Phi^{\mu}+\widetilde{\Gamma}^{\mu}_{
\nu\rho}\partial_{t}\Phi^{\nu}\partial_{t}\Phi^{\rho} \\ \nonumber
&=& D^{2}_{t}\Phi^{\mu}+G^{\mu
i}a_{i}|\partial_{t}\Phi|^{2}-G^{\mu\nu}\frac{\partial
f(\gamma)}{\partial\gamma}|\partial_{t}\Phi|^{2}\partial_{\nu}\gamma
\\  &-&\frac{\partial
f(\gamma)}{\partial\gamma}\partial_{t}\gamma\partial_{t}\Phi^{\mu}-2a_{i}\partial_{t}\Phi^{i}\partial_{t}\Phi^{\mu}.
\end{eqnarray}
where $\widetilde{\Gamma}^{\mu}_{ \nu\rho}$ is the Lev-Civitra
connection for the augmented target space metric $\widetilde{G}$ .
When $f(\gamma)$ satisfies the condition

\beq f(t)=6\beta(t),\eeq or, equivalently,

\beq \frac{\partial f(\gamma)}{\partial
\gamma}=6\sqrt{A}(\partial_{t}\gamma)^{n-1},\eeq we find that the
equation

\beq \widetilde{D}^{2}_{t}\Phi^{\mu}=\frac{1}{2}\frac{\partial
f(\gamma)}{\partial\gamma}\partial_{t}\gamma\partial_{t}\Phi^{\mu}-2a_{i}\partial_{t}\Phi^{i}\partial_{t}\Phi^{\mu}
\eeq is equivalent to Eqs. (17) and (18) with respect to $\mu=0$
and $\mu\neq0$ respectively. Eq. (24) is the equation for a
geodesic in a non-ffine parametrization. In terms of the new time
variable $\widetilde{t}$ defined by

\beq d\widetilde{t}=2\Omega^2e^{-f(\gamma)/2}dt,\eeq we have

\beq \widetilde{D}_{\widetilde{t}}^{2}\Phi^{\mu}=0. \eeq This is
the equation of an affinely parameterized geodesic with respect to
the conformally-rescaled metric. The Friedmann equation now can be
written as

\begin{equation}
\widetilde{G}_{\mu\nu}\partial_{\widetilde{t}}\Phi^{\mu}\partial_{\widetilde{t}}\Phi^{\nu}=\cases{{-signV}
 & $V\ne0$ \cr {0}\ \,
 & $V=0$\,.}
\end{equation}
Since null geodesics are unaffected by the conformal rescaling,
the cosmological trajectories with $V=0$ is null geodesic with
respect to $G$ as well. What is interesting is that it also shows
that in flat space the the cosmological trajectories driven only
by kinetic term are also geodesic with respect with with another
choice of metric discussed below .Whereas, for $V\neq0$, the
cosmological trajectories in the augmented target space are
geodesics only with respect to a class of conformally-rescaled
metrics which are restricted by the form of $f(\gamma)$.
Especially, to the GR case, that is, $n=1$, $f$ have the form of
$6\sqrt{A}\gamma$ which is equivalent to what is discussed
in\cite{gr}.

\section{Purely Kinetic Energy Driving Cosmologies }

In this case the motion of scalar is discussed in an augmented
space with  different coordinates and  metric. Define $(N+1)$
variables

\beq \Theta^{\mu}=(\beta, \phi^{\alpha})\eeq as the coordinates.
We take the metric as

\beq
\widehat{G}_{\mu\nu}d\Theta^{\mu}d\Theta^{\nu}=-e^{-6(n-1)\beta}d\beta^{2}+G_{\alpha\beta}d\phi_{i}d\phi^{j}.
\eeq In this notation, The equation of motion becomes

\beq
\widehat{D}^{2}_{t}\phi^{i}+3\partial_{t}\beta\partial_{t}\phi^{i
}=0,\eeq and Friedmann equation for $V=0$ becomes

\beq
A|\partial_{t}\phi|^{2n}-(\partial_{t}\beta)^{2}=ke^{-2\beta}.\eeq
From the Friedmann equation one can know that the  cosmological
trajectories do not have  global consistent causal property. An
exception is $n=1$, the GR case\cite{gr}. From Eq. (31) for $n=1$,
the curve is null, timelike and spacelike with respect to
$k=0$,$k=-1$, and $k=1$ respectively. For an arbitrary n, using
Friedmann equation for $V=0$, the acceleration equation can be
written as

\beq
\widehat{D}^{2}_{t}\beta+3(\partial_{t}\beta)^{2}=-(3n-1)ke^{-2\beta}.\eeq
Combined with the Eq.(30), we get a single equation

\beq
\widehat{D}^{2}_{t}\Theta^{\mu}+3(\partial_{t}\beta)\partial_{t}\Theta^{\mu}=(3n-1)ke^{-2\beta-6(n-1)\beta}\widehat{D}^{\mu\nu}\partial_{\nu}\gamma.\eeq
Introducing another time-coordinate

\beq d\widehat{t}=e^{-3\beta}dt, \eeq One can obtain

\beq
\widehat{D}_{\widehat{t}}^{2}=(3n-1)ke^{4\beta-6(n-1)\beta}\widehat{G}^{\mu\nu}\partial_{\nu}\beta.\eeq
For $k=0$, this is the  equation of an null affinely parameterized
geodesic. For $k\neq0$, the cosmological trajectory in the
augmented target space is not a geodesic for the right-hand-side
of Eq.(31) does not vanish. However, we will find that the
trajectory can be viewed as the projection of a geodesic on a
hypersurface in a `doubly-augmented' target space of dimension
(N+2) that is foliated by hypersurfaces isometric to the augmented
target space.

Consider the (N+2) variables

\beq \Xi^{A}=(\Theta^{\mu},\Xi^{*})\eeq as maps from the
cosmological trajectory to the `doubly-augmented' target space. We
adopt the metric

\beq
\widehat{G}_{AB}d\Xi^{A}d\Xi^{B}=\widehat{G}_{\mu\nu}d\Theta^{\mu}d\Theta^{\nu}+
\widehat{G}_{**}(d\Theta^{*})^{2}, \eeq where

\beq \widehat{G}_{**}=\frac{3n-1}{3n-5}ke^{(6n-10)\beta},\eeq We
need another equation to constrain the motion of the additional
variable $\Xi^{*}$. Take the motion equation of $\Xi^{*}$ as

\beq
\partial_{\widehat{t}}\Theta^{*}=e^{(10-6n)\beta},
\eeq
 With this choice,
one can obtain

\beq \widehat{D}_{\widehat{t}}^{2}\Xi^{A}=0. \eeq

Thus, all cosmological trajectories driven by kinetic term is
geodesics in the doubly-augmented target space. The cosmological
trajectories in augmented target space are corresponding to  the
projection of these geodesics on a motive hypersurface whose
motion is restricted by equation (39). However, for $k=0$, the
cosmological trajectories are always geodesics not only in
doubly-augmented target space but also in augmented target as
discussed in Sec. 2.

One can easily find that to the case of $n=1$, the two classes of
coordinates and metric in Sec. 2 and this Sec. are equivalent. It
is the different effective Friedmann equations that lead to the
existence of different choices to set up augment target space.

\section{Acceleration Cone}

From Eq.(8), it is shown that only when potential $V>0$ could
acceleration occur. In this section we only discuss the case of
flat universe. In the notation of Sec.2, using Friedmann equation
to eliminate $2V$ one can rewrite the acceleration equation as

\beq
\partial_{t}^{2}S=3nAS\rho^{n-1}[\frac{(\partial_{t}\gamma)^{2}}{3n}-|\partial_{t}\phi|^{2}].
\eeq We will see that the acceleration acquires a geometrical
meaning in the augmented target space with the following
acceleration metric

\beq
G^{acc}_{\mu\nu}d\Phi^{\mu}d\Phi^{\nu}=-\frac{1}{3n}d\gamma^{2}+G_{ij}d\phi^{i}d\phi^{j}.
\eeq One finds that

\beq
G^{acc}_{\mu\nu}\partial_{t}\Phi^{\mu}\partial_{t}\Phi^{\nu}=-\frac{1}{3n}(\partial_{t}\gamma)^{2}+|\partial_{t}\phi|^{2}
\eeq Substituting it into Eq. (8) we obtain

\beq
\partial_{t}^{2}S=-3nAs\rho^{n-1}G^{acc}_{\mu\nu}\partial_{t}\Phi^{\mu}\partial_{t}\Phi^{\nu}
\eeq

Thus,we come to the conclusion that a universe is accelerating
when the tangent to its trajectory lies within a subcone of the
lightcone defined by the acceleration metric on the augmented
target space.

\section{Conclusions}
In this paper, we have presented an geometric analysis to classify
the homogeneous isotropic cosmology in the context of effective
Friedmann equation in multi-scalar models with an arbitrary scalar
potential $V$. Two models of interest is discussed respectively.
The first is analysis in the flat universe. In this case, the
target space of $N$ scalar fields $\phi$ is augmented to a larger
$(N+1)$-dimensional Lorentzian signature sapce,where function of
the scalar factor plays the role of time. It is found that when
$V=0$, flat cosmological trajectories correspond to null geodesic
in the augmented target space. But for $V\neq0$, the cosmological
trajectories are geodesic only when a class of conformally
rescaled metrics are chosen. The conformal factor greatly depend
on the form of potential $V$.

The other is the universe driven by purely kinetic energy. A
distinguished simple time-coordinate is chosen at the price of the
relatively complicated metric. In the augmented space we take the
logarithm of the scale factor as the time-coordinate. In this
case, for flat universe, same conclusion is reached as the former
case. While for $k\neq0$, the cosmological trajectories are no
longer geodesics. However, they can be viewed as geodesics in a
doubly-augmented target space with dimension $(N+2)$  Thus, all
purely kinetic energy driving cosmological trajectories can be
viewed as projection of these geodesics.

Though the choice of time-coordinate is different in these two
case, the geometric properties for flat universe driven by purely
kinetic energy is same. What is more, when $n=1$, these two
augmented target manifolds are equivalent.

%%%%%%%%%%%%%%%%%%%%%%%%%%%%%%%%%%%%%%%%%%%%%%%%%%%%%%%%%%%%%%%%%%%%%%%%%%%%%%%%%%%

\section*{Acknowledgments}
Qi Guo thanks Prof. Paul Townsend for discussion on the paper
\cite{gr}. Also she thanks Hong-Sheng Zhang, Zong-Kuan Guo, Hao
Wei, Hui Li and Da-Wei Pang for helpful discussion.This project
was in part supported by Chinese Academy of Sciences,  by NNSFC
under Grant No.10325525 and No.90403029, and also by MSTC under
Grant No.TG1999075401.


\begin{thebibliography}{00}

\bibitem{observation}
  V.Sahni and A.A. Starobinsky,
  Int. J. Mod.Phys. D  {\bf 9}, 373 (2000)
  [arXiv:astro-ph/9904398]; T. Padmanabhan, Phys. Rept. {\bf
  380}, 235 (2003) [arXiv:astro-ph/0212290].
\bibitem{que}
R.R. Caldwell, R. Dave and P. J. Steinhardt, Phys. Rev. Lett. {\bf
80}, 1582 (1998) [astro-ph/9708069]; C. Wetterich, Nucl. Phys. B
{\bf 302}, 668 (1998) P. J. E. Peebles and B. Ratra, Astrophys. J.
{\bf 325}, L17 (1988) P. J. Steinhardt, L. M. Wang and I. Zlatev,
Phy. Rev. D {\bf 59} 123504 (1999) [astro-ph/9812313].
\bibitem{kess}
 T. Chiba, T. Okabe and M. Yamaguchi, Phys. Rev. D {\bf 62},
 023511 (2000);
 C. Armendariz-Picon, V. Mukhanov, and P. J. Steinhardt, Phys.
Rev. Lett. {\bf 85}, 4438 (2000);
 C. Armendariz-Picon and V. Mukhanov, Phys. Rev. D {\bf 63}, 103510
 (2000).
\bibitem{kkess}
Robert J. Scherrer, Phys.Rev.Lett. {\bf 93}, 011301 (2004)
\bibitem{phantom}
R. R. Caldwell, Phys. Lett. B 545, 23 (2002) [astro-ph/9908168];
R. R. Caldwell, M. Kamionkowski and N. N. Weinberg, Phys. Rev.
Lett. 91, 071301 (2003) [astro-ph/0302506]. S. M. Carroll, M.
Hoffman and M. Trodden, Phys. Rev. D 68, 023509 (2003)
[astro-ph/0301273].
\bibitem{doublescalar}
Hao Wei and Rong-Gen Cai [hep-th/0501160]; Zong-Kuan Guo, Rong-Gen
Cai, Yuan-Zhong Zhang [astro-ph/0412624];W. Hu,
[astro-ph/0410680]; B. Feng, X. L. Wang and X. M. Zhang,
[astro-ph/0404224].
\bibitem{nogo}
A. Vikman, astro-ph/0407107.
\bibitem{gr}
Paul K. Townsend and Mattias N.R. Wohlfarth, Class.Quant.Grav.
{\bf 21} 5375 (2004)
\bibitem{effF}
Edmund J. Copeland, Seung-Joo Lee, James E. Lidsey and Shuntaro
Mizuno,  Phys.Rev. D {\bf 71},  023526 (2005).
\bibitem{effF2}
Shinji Tsujikawa and M. Sami, Phys.Lett. B {\bf 603} 113-123
(2004).
\bibitem{GB}
C. Charmousis and J. Dufaux, Class. Quant. Grav. {\bf 19}, 4671
(2002);C. Germani and C. F. Sopuerta, Phys. Rev. Lett. {\bf 88},
231101 (2002).
\bibitem{RS1}
Randall L, Sundrum R 1999 Phys.Rev.Lett. {\bf 83} 3370; Randall L,
Sundrum R 1999 Phys.Rev.Lett. {\bf 83} 4690; Philippe Brax and
Carsten van de Bruck, Class.Quant.Grav. {\bf 20}, R201-R232
(2003).



\end{thebibliography}
\end{document}